\documentclass[aps,prl,twocolumn,floats,showpacs,superscriptaddress]{revtex4}
\usepackage{graphicx,epsfig}
\usepackage{times}
\usepackage{graphics,dcolumn,bm,float}
\usepackage{amssymb,amsmath,rotate,color}
\usepackage{psfrag}

\newcommand{\be}{\begin{equation}}
\newcommand{\ee}{\end{equation}}
\newcommand{\bearr}{\begin{eqnarray}}
\newcommand{\eearr}{\end{eqnarray}}

\begin{document}
\unitlength 1 cm

\title{Variational RVB wave function for the spin-1/2 Heisenberg Model on honeycomb lattice }
\author
{Zahra Nourbakhsh{\footnote {Electronic address:
z.nourbakhsh@ph.iut.ac.ir}}}

\author{Farhad Shahbazi{\footnote {Electronic address:
shahbazi@cc.iut.ac.ir}}}

\author{S. A. Jafari{\footnote {Electronic address:
sa.jafari@cc.iut.ac.ir}}}

\affiliation {\it Dept. of Physics , Isfahan University of
Technology, 84156-83111, Isfahan, Iran}

\author{G. Baskaran}
\affiliation{\it Institute of Mathematical Sciences, Chennai 600113, India}

\date{\today}

\begin{abstract}
In this work, a long-range resonating valence bond state is proposed
as a variational wave function for the ground state of the $S=1/2$
antiferromagnetic Heisenberg model on the honeycomb lattice. Employing Variational
Monte Carlo (VMC) method, we show that the ground state energy
obtained from such RVB wave function, lies well below the energy
of the N\'eel state and  compares very well to the energies evaluated
from spin-wave theory and series expansion method. We also obtain the spin-spin correlation
function along zig-zag and armchair directions and find that the two
correlations are different, which indicates the anisotropic nature
of the system. We compare our results with the square lattice  and
we show that although the quantum fluctuations on honeycomb lattice
are much stronger, but do completely not destroy the magnetic order.
\end {abstract}



\maketitle  

{\em Introduction}: The resonating valence bond (RVB) state was
originally proposed
 by Anderson and Fazekas as a possible ground state for the
$S=1/2$ Heisenberg spins with nearest neighbor  anti-ferromagnetic coupling
on triangular lattice~\cite{anderson73,fazekas74}. They found that
for  anisotropic Heisenberg model, near the Ising limit, the
liquid-like RVB state is energetically more favorable
than the N{\'e}el state. In such a state the $S=1/2$ atoms
residing on the lattice points, form singlet valence bonds in
pairs and so lose some of their anti-ferromagnetic exchange energy
with respect to the N{\'e}el order. In order to regain  some of the
lost exchange energy, they have to resonate   quantum mechanically
among many different pairing configurations. These states form the
basis of Pauling's early theories of aromatic molecules such as
benzene~\cite{Pauling},  however his theory was unable to give a proper description of
the metallic state. The idea that the RVB state of spin-liquid type
may give a precise picture of the two dimensional quantum
anti-ferromagnet was once the most attractive topic after the
discovery of the high-$T_c$  superconductivity, when
Anderson~\cite{anderson87} suggested that an RVB state naturally leads to
incipient superconductivity from preformed singlet pairs in the
parent insulating state.

The anti-ferromagnetic Heisenberg model arises naturally in a Mott
insulator, in which a system with an odd number of electrons per
unit cell is insulating due to the strong Coulomb repulsion
between two electrons on the same site ($U$). In such systems the
kinetic exchange mechanism due to virtual hopping  between
anti-parallel spin configurations  leads to anti-ferromagntic
exchange $J=4t^2/U$ between the spins, where $t$ is the hopping
integral~\cite{anderson59}. For $S=1/2$  and in low dimensions,
strong quantum fluctuations make the RVB liquid more favorable
than the classically ordered N{\'e}el state.  RVB is  a fair
description of anti-ferromagnetic Heisenberg linear  chain, where
there is no long-range order according to the exact Bethe's
solution~\cite{bethe31}. The $S=1/2$ anti-ferromagnet Heisenberg
model on square lattice has been extensively studied~\cite{rmp}
and it has been established that there is a N{\'e}el order in the
system, although the quantum fluctuations  reduce the staggered
magnetization  with respect to  its   classical value.

Apart from the fabrication of Graphene sheets which brought
the honeycomb structure to the focus of condensed matter community, the
recent discovery of compounds such as
InCu$_{2/3}$V$_{1/3}$O$_3$~\cite{Kataeve} in which the  Cu$^{+2}$ ions in
the copper-oxide layers form a two-dimensional $S=1/2$
anti-ferromagnt Heisenberg on a honeycomb lattice are our
motivations for this study. Also the recent progress in the field
of ultracold atoms and trapping techniques~\cite{Aubin} along with
the ability to tune the interaction parameters via the Feshbach
resonance~\cite{Feshbach} can be thought of another way to realize
Heisenberg spins (of localized fermions) on a honeycomb optical lattice.

In this work, we study the $S=1/2$ Heisenberg spins with
nearest-neighbor anti-ferromagnetic interactions on the honeycomb
lattice.  The number of nearest neighbors in
honeycomb lattice equals 3, which is less than 4 of the square lattice, leading to
enhancement of quantum fluctuations. This suggests that the RVB state could be a
better variational wave function
for honeycomb than the square lattice. In this work we choose the
variational RVB wave function proposed by Liang, Doucot and Anderson in
the context of HTSC cuprates~\cite{LDA} and show that RVB ansatz in honeycomb
lattice gives very good results for the ground state energy by comparing with
other methods.

\begin{figure}[ht]
  \begin{center}
    \vspace{-1.5cm}
    \includegraphics[scale=0.35]{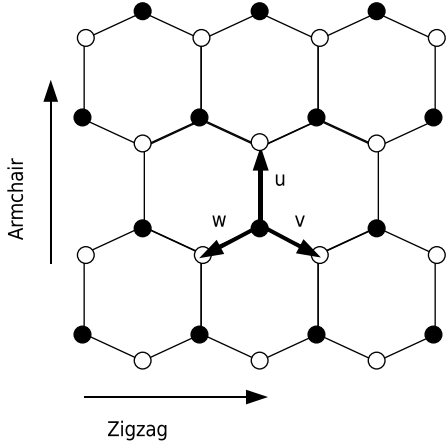}
    \vspace{-2.0cm}
    \caption{Honeycomb lattice as a superposition of two simple
    monoclinic Bravais lattices denoted by filled and empty circles.}
    \label{honey}
  \end{center}
\end{figure}

{\em Numerical calculation}:
Consider an system of atoms with $S=1/2$ on a a honeycomb lattice,
consisting of two interpenetrating bravais sublattices~(Fig.
\ref{honey}). The Heisenberg Hamiltonian with anti-ferromagnetic
nearest neighbor interactions for this system is given by:
\begin{equation}\label{hamiltonian}
   H=J\sum_{\langle i,j\rangle} {\mathbf S}_i.{\mathbf S}_j,
\end{equation}
where $J>0$ and $\langle i,j\rangle$ denotes the nearest neighbors. Since the
ground state of such Hamiltonian is a spin singlet~\cite{marshall},
we employ a RVB state as trial wave function. After its optimization, we calculate
the ground state properties. The RVB trial wave function can be
considered as summation over the all possible singlet pairings
between each spin in one sublattice (say $A$) to spins on the
other sublattice ($B$).
Therefore the trial wave function is given by:
\begin{eqnarray}
   \label{wave}
   |\Psi\rangle &=&\sum_{{i_\alpha\in A} \atop {j_\beta\in B}}
   h(i_1-j_1)...h(i_n-j_n) (i_1,j_1)...(i_n,j_n)\nonumber\\
   &=&\sum_i w(c_i)|c_i\rangle
\end{eqnarray}
Where $(i,j)$ represent a singlet bond.
Denoting spin up state ($|\uparrow\rangle$) by $\alpha$  and
spin down state ($|\downarrow\rangle$) by $\beta$, it can be expressed as:
\begin{equation}
   (i,j)=\frac{1}{\sqrt2}(\alpha_i\beta_j-\beta_i\alpha_j).
\end{equation}
In Eq.~(\ref{wave}), $|c\rangle$ stands for a valence bond configuration which
can be represented by:
\begin{equation}
   |c\rangle=\prod_{a=1}^{n}(i_a,j_a).
\end{equation}
The weight of a configuration $|c\rangle$ of valence bonds is given by
\begin{equation}
   w(c)=\prod_{a=1}^{n}h(i_a-j_a),
\end{equation}
where $h$ is a pairing function describing a singlet bond as a
function of bond length.

\begin{figure}[t]
  \begin{center}
    \vspace{0.0 cm}
    \vspace{-1.5cm}
    \includegraphics[scale=0.35]{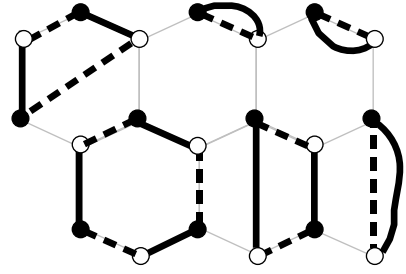}
    \vspace{-4.0cm}
    \caption{The loop covering $\langle c_1|c_2\rangle$ is the superposition of $|c_1\rangle$ (solid
line) and $|c_2\rangle$ (dashed line) on a honeycomb lattice.
$|c_1\rangle$ and $|c_2\rangle$ are two singlet valence bond
configurations with equal weight factors. In this example, there
are six loops.}
    \label{loops}
  \end{center}
\end{figure}

This wave function contains the two limiting cases of the
nearest-neighbor RVB liquid  (when $h(l)=0$ for $l>1$) and the
N{\'e}el state (for which $h(l)$ is independent of the bond
length).  Sutherland derived a set of simple rules for estimating $\langle
c_1|{\mathbf S}_i.{\mathbf S}_j|c_2\rangle$~\cite{sutherland}. It is  easy to show
that the overlap between two configurations $|c_1\rangle$ and
$|c_2\rangle$ is given by $\langle c_1|c_2\rangle=2^{N(c_1,c_2)}$,
where $N(c_1,c_2)$ is the number of loops  in the overlap of the
two configurations (Fig. \ref{loops}). To compute the ground
state energy as well as the spin-spin correlation functions, we
use the following rules for the matrix elements of the Hamiltonian:
(i) $\langle c_1|{\mathbf S}_i.{\mathbf S}_j|c_2\rangle=0$, if $i$ and $j$
belong to two different loops;
(ii) $\langle c_1|{\mathbf S}_i.{\mathbf S}_j|c_2\rangle=\pm \frac{3}{4}\langle
c_1|c_2\rangle$,  if $i$ and $j$ belong to the same loop, with a
minus sign when $i$ and $j$ belong to
 two different sublattices, and a plus sign otherwise.

Using the wave function~(\ref {wave}) along with the
Hamiltonian~(\ref {hamiltonian}), the ground state energy is given by:
\begin{eqnarray}\label{gs}
E&=&\frac{\langle\Psi|H|\Psi\rangle}{\langle\Psi|\Psi\rangle}\nonumber\\
&=&\sum_{c_1,c_2}\frac{w(c_1) w(c_2)
\langle c_1|c_2\rangle}{\langle\Psi|\Psi\rangle}\times\frac{\langle c_1|H|c_2\rangle}{\langle c_1|c_2\rangle}\nonumber\\
&=&\sum_{c_1,c_2}P(c_1,c_2)\times E(c_1,c_2),
\end{eqnarray}
in which $E(c_1,c_2)$ is the contribution in the ground states
energy  arising from the overlap of configurations $|c_1\rangle$
and $|c_2\rangle$ with the weight $P(c_1,c_2)=\frac{w(c_1) w(c_2)
\langle c_1|c_2\rangle}{\langle\Psi|\Psi\rangle}$. Such overlaps
can be graphically represented by loop coverings depicted in
Fig.~\ref{loops}~\cite{FradkinBook}. The number of terms in this
expression  diverges exponentially with the system size, making
the exact evaluation of the summation impossible. To overcome
this difficulty, we use Monte Carlo approach based on important
sampling for evaluating the ground state energy and also the
spin-spin correlations. since  $P(c_1,c_2)$ is positive, here
there is no minus sign problem, and the Monte Carlo estimates can
be made very precise.

  Consider two configurations $c_1$ and
$c_2$ with a given loop coverage. Using the standard Metropolis
algorithm, loop configurations are updated by randomly choosing a
couple of sites  and exchanging their end point  connections with
a probability that satisfies the detailed balance
condition~\cite{LDA}. The matrix elements are evaluated according
to Sutherland's rule, once the loop covering associated with the
two configurations is known. In our calculations, boundary
conditions are take to be periodic. The equilibrium state does
not depend on the initial state, but in order to reach the
equilibrium distribution faster, we started with a dimer state.
In following we discuss our results for short-range and
long-range RVB states.

\begin{figure}[t]
   \begin{center}
   \includegraphics[scale=0.55,angle=-90]{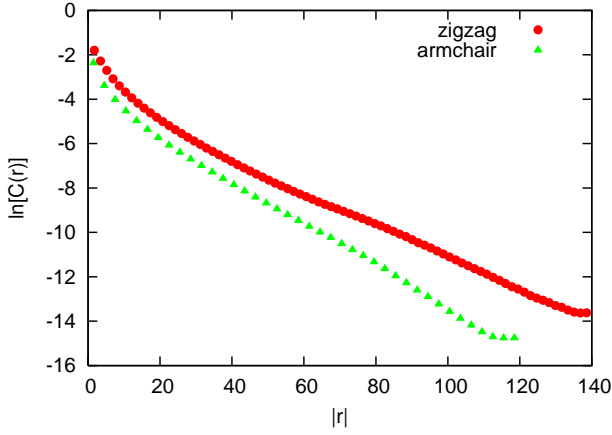}
   \end{center}
   \caption{ The spin-spin correlation  as a function of distance for the state labeled
    $(1,3,5)$. See the text for explanations of the state.
    Circles denote zigzag direction data and triangles  denote armchair data.}
   \label{fshort}
\end{figure}

{\em Short-range RVB wave function}: First we consider some wave
functions with short singlet bonds. One example of this type is
the nearest-neighbor RVB (NNRVB) trial wave functions for which
$h(1)=1$ and $h(l)=0$ for larger distances $l>1$. We choose $l$
to be the Manhatan distance defined for two points separated by a
vector $n_u {\mathbf u}+n_v {\mathbf v}$ in Fig.~\ref{honey} by
$l=|n_u|+|n_v|$.  Defining $a_l=\frac{h(2l+1)}{h(2l-1)}$, we
investigate other examples such as  $(1,3)$ state with one
variational parameter $a_{1}=\frac{h(3)}{h(1)}$ and $a_{l}=0$ for
$l>2$; $(1,3,5)$ state with two variational parameters
$a_{1}=\frac{h(3)}{h(1)}$ $a_{2}=\frac{h(5)}{h(3)}$; and
exponential state for which $h(l)$ decays exponentially for large
distances. In this state we have three variational parameters
$a_1,a_2$ and $a_l=\mbox{const. for } l\geq2$. Among the short
range states, the exponential state has lowest energy (see
Table~\ref{tshort}). The statistical error in evaluation of
energy from Monte Carlo calculation is $0.0002$ and we check that
the finite-size effects  are less than this error.  We have also
computed the spin-spin correlation function defined as:
\begin{equation}
  C_{ij}=\sum_{c_{1},c_{2}}P(c_{1},c_{2})
  \frac{\langle c_1|{\mathbf S}_{i}.{\mathbf S}_{j}|c_2\rangle}{\langle c_1|c_2\rangle},
\end{equation}
which results in exponentially decaying of the correlation  at
large distances, indicating the absence of  long- range order for
this variational states (Fig.~\ref{fshort}). As it can be seen in
Fig.~\ref{fshort}, the correlation length along  zigzag direction is
larger  than armchair direction. The correlation lengths $\xi$ in units
of the lattice spacing are listed in Table~\ref{tshort}.

\begin{table}[h]
\begin{tabular}{cccc}
\hline
State&$\frac{-E_{\circ}}{J}$&$\xi$&$h(l)$\\
\hline
NNRVB&$0.4310(2)$&$\xi_a=0.84$&$h(1)=1$\\
$$&$$&$\xi_z=1$&$h(l)=0,~l>1$\\
$(1,~3)$&$0.5087(2)$&$\xi_a=4.7$&$a_1=\frac{2}{9}$\\
$$&$$&$\xi_z=5.9$&$a_l=0,~l>1$\\
$(1,~3,~5)$&$0.5319(2)$&$\xi_a=9.55$&$(a_1,~a_2)=(\frac{1}{9},\frac{2}{3})$\\
$$&$$&$\xi_z=11.1$&$h(l)=0,~l>2$\\
Exponential&$0.5437(2)$&$-$&$(a_1,~a_2)=(\frac{2}{21},\frac{1}{4})$\\
$$&$$&$-$&$a_l=0.32,~l>3 $\\
\hline
\end{tabular}
\caption{Ground state energy, spin-spin correlation length and optimized parameters for  the short ranged RVB states. The subscripts $a,z$ stand for armchair and zigzag directs,
respectively.}
\label{tshort}
\end{table}
\begin{figure}[t]
  \begin{center}
     \includegraphics[scale=0.55,angle=-90]{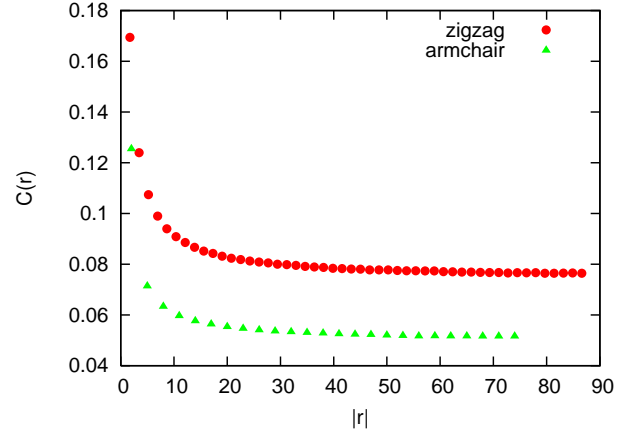}
     \label{flong}
  \end{center}
\caption{spin-spin correlation as a function of
distance for the states with algebraically decaying states with $p=3$ (see
the text), versus $r$  along the zigzag (circle) and armchair
(triangle) directions.}
\end{figure}

\begin{table}[hb]
\begin{tabular}{cccc}
\hline
 $p$&$\frac{-E_{\circ}}{J}$&$M_s$&$(a_1,a_2)$\\
 \hline
$2$ & $0.5434(2)$ & $0.30(2)$ & $(\frac{1}{16},\frac{2}{9})$\\
$2.5$ & $0.5440(2)$ & $0.26(2)$ & $(\frac{2}{23},\frac{2}{9})$\\
$3$ & $0.5438(2)$ & $0.25(2)$ & $(\frac{2}{19},\frac{1}{4})$\\
$3.5$ & $0.5431(2)$ & $0.20(2)$ & $(\frac{1}{9},\frac{1}{4})$\\
$4$ & $0.5430(2)$ & $0.19(2)$ & $(\frac{1}{8},\frac{2}{7})$\\
\hline
\end{tabular}
\caption{Ground state energy, staggered magnetization  and optimized parameters for  the long ranged RVB states.}
\label{tlong}
\end{table}

{\em Long-range wave functions}: Next we study spin-spin
correlations and energy for the class of wave functions which
behave like $h(l)\propto l^{-p}$ in long distances. To optimize the energy,
we choose $a_1=\frac{h(3)}{h(1)}$,  $a_3=\frac{h(5)}{h(3)}$ as
variational parameters and optimize them for various exponents $p$
defined by $h(l)=h(5)(5/l)^p$ for $l>5$. For long-bond singlets
the spin-spin correlation  function does not vanish at large
distances and so the staggered magnetization can be determined from
the tails of the correlation function. Because of the anisotropic
nature of the  honeycomb lattice, the correlations along zigzag
and armchair directions are different hence  we define the
magnetization by
 \begin{equation}
 M^s=\sqrt{\lim_{r\rightarrow
 L/2}\frac{C_{zigzag}(r)+C_{armchair}(r)}{2}},
 \end{equation}
where $L$ is the linear size of the system.

 The optimal values for the parameters as well as ground state energies and staggered
magnetizations within this class of states are listed in Table~\ref{tlong}
for various values $p=2, 2.5, 3, 3.5$ and $4$. For this set
of exponents, we found that energy is minimum for $p=2.5$ with
$E_{\circ}=-0.5440J\pm0.0002J$, which is lower than the energy of
the short ranged RVB states. The staggered magnetization for $p=2, 2.5, 3,
3.5$ and $4$ are $\%60, \%52, \%50, \%40,$ and $\%38$ of N{\'e}el
state, respectively. The magnetization decreases when $p$
increases and seems to disappear when $p>4$.

{\em Conclusion}: In summary, we presented a variational Monte Carlo estimate of
the ground state energy for the $S=1/2$ AF Heisenberg model on honeycomb
lattice employing  both short-range and long-range RVB trial  wave functions.
Similar to square lattice~\cite{LDA}, we found that the long-bond wave functions give a better description of the ground state in honeycomb lattice.
The optimized values of ground state energy and staggered magnetization are
 $E_{\circ}/J=-0.5440\pm0.0002$ and  $M^s=0.26$, respectively. The ratio of the RVB ground state energy to the N{\'e}el state energy
is $\frac{E_{RVB}}{E_{N}}=\frac{0.5540}{0.375}=1.48$, so RVB
energy is \%48 less than the N{\'e}el state. In square lattice,
this ratio is \%34, which can be justified in terms of stronger
quantum fluctuation in honeycomb lattice due to its smaller
coordination number. The quantum fluctuations also  reduce the
order parameter by about $\%50$ with respect to classical
N{\'e}el order.

Our results are in excellent agreement  with other numerical
methods, such as the series expansion~\cite{series} which gives
the ground state energy $\frac{E_{\circ}}{J}=-0.5443\pm0.0003$
and magnetization $M^s=0.266\pm0.009$ as well as   spin wave
calculation in first approximation with
$\frac{E_{\circ}}{J}=-0.5324$ and $M^s=0.2418$ and in second
approximation with $\frac{E_{\circ}}{J}=-0.5489$ and
$M^s=0.2418$~\cite{sw}. This shows that RVB picture captures the
main properties of the ground state for the AF Heisenberg model
on the honeycomb lattice with higher precision than square
lattice, leading to the conclusion that the stronger are quantum
fluctuations, the more favorable is the RVB ground state.
Therefore it seems that the RVB state is a good candidate for the
ground state of the AF Heisenberg model on the honeycomb lattice.
One can take a long ranged RVB as a reference starting point to
proceed with the calculations of excitation
energies~\cite{Mosadeq}. Including charge fluctuations in such an
RVB state by adding hopping of underlying electrons has also been
investigated~\cite{jafari-dmft-honeycomb,VijayBaskaran}. This
view point can be taken as a possible rout to describe the Dirac
liquid semi-metallic in terms of RVB wave
functions~\cite{BaskaranJafari}.

{\em Acknowledgments:}
 F.S. and S.A.J. thank the Abdus Salam ICTP for the hospitality during a
short term summer visit, where some part of this work was done.
S.A.J. was supported by ALAVI group Ltd. We thank H. Mosadeq for
useful discussions.

\end{document}